\batchmode

\documentclass[aps,prl,reprint,superscriptaddress,showpacs,longbibliography,floatfix,footnoteinbib]{revtex4-1}
\usepackage[latin9]{inputenc}
\usepackage{color}
\usepackage{amsmath}
\usepackage{amssymb}
\usepackage{graphicx}
\usepackage[pdftex,
 bookmarks=false,
 breaklinks=false,pdfborder={0 0 1},backref=false,colorlinks=true]
 {hyperref}
\hypersetup{pdftitle={Title},
 plainpages=false,pdfpagelabels,linkcolor=blue,urlcolor=blue,citecolor=blue,pdfdisplaydoctitle=true,pdfduplex=DuplexFlipLongEdge}

\makeatletter
\usepackage{units}\usepackage{wasysym}

\definecolor{orange}{rgb}{0.50, 0.20, 0.0}

\usepackage{bm}
\usepackage{braket}

\makeatother

\begin{document}
\title{Fractal Quasicondensation in One Dimension}
\author{Flavio Riche}
\affiliation{CeFEMA, LaPMET, Instituto Superior T\'ecnico, Universidade de Lisboa,
Av. Rovisco Pais, 1049-001 Lisboa, Portugal}
\author{Miguel Gon\c{c}alves}
\affiliation{Princeton Center for Theoretical Science, Princeton University, Princeton,
NJ 08544}
\author{Bruno Amorim}
\affiliation{Centro de F\'isica das Universidades do Minho e Porto, LaPMET, Universidade
do Minho, Campus of Gualtar, 4710-057, Braga, Portugal}
\affiliation{International Iberian Nanotechnology Laboratory (INL), Av. Mestre
Jos\'e Veiga, 4715-330 Braga, Portugal}
\author{Eduardo V. Castro}
\affiliation{Centro de F\' isica das Universidades do Minho e Porto, LaPMET, Departamento
de F\'isica e Astronomia, Faculdade de Ci\^encias, Universidade do Porto,
4169-007 Porto, Portugal}
\affiliation{Beijing Computational Science Research Center, Beijing 100084, China}
\author{Pedro Ribeiro}
\affiliation{CeFEMA, LaPMET, Instituto Superior T\'ecnico, Universidade de Lisboa,
Av. Rovisco Pais, 1049-001 Lisboa, Portugal}
\affiliation{Beijing Computational Science Research Center, Beijing 100084, China}
\begin{abstract}
We unveil a novel mechanism for quasicondensation of hard-core bosons
in the presence of quasiperiodicity-induced multifractal single-particle
states. The new critical state, here dubbed fractal quasicondensate,
is characterized by natural orbitals with multifractal properties
and by an occupancy of the lowest natural orbital, $\lambda_{0}\simeq L^{\gamma}$,
which grows with system size but with a non-universal scaling exponent,
$\gamma<1/2$. In contrast to fractal quasicondensates obtained when
the chemical potential lies in a region of multifractal single-particle
states, placing the chemical potential in regions of localized or
delocalized states yields, respectively, no condensation or the usual
1D quasicondensation with $\gamma=1/2$. Our findings are established
by studying one-dimensional hard-core bosons subjected to various
quasiperiodic potentials, including the well-known Aubry-Andr\'e model,
employing a mapping to non-interacting fermions that allows for numerically
exact results. We discuss how to test our findings in state-of-the-art
ultracold atom experiments.
\end{abstract}
\maketitle

\section{Introduction}

The localization of single-particle wave functions predicted by Anderson
\cite{anderson1958absence} can be induced by uncorrelated disorder
or by quasiperiodic (QP) perturbations incommensurate with the underlying
crystal. Quasiperiodicity can induce localization-delocalization transitions
even in one dimension \cite{aubry1980analyticity}, where any finite
amount of uncorrelated disorder immediately localizes the wave function
\cite{PhysRevLett.42.673,PhysRevLett.47.1546,lee1985disordered,continentino2017quantum}.
QP modulations may also lead to critical states with multifractal
properties \cite{liu2021anomalous,ganeshan2015nearest,gonccalves2021hidden,gonccalves2022exact,gonccalves2020disorder}.
Such critical states arise at localization phase transitions and were
also shown to ensue in extended areas of the phase space, where they
can coexist with localized and extended states, albeit separated by
the so-called mobility edges into different spectral regions \cite{ribeiro2013strongly,wang2020one,biddle2011localization,deng2019one,gonccalves2023critical}.

Interest in QP single-particle systems, kickstarted in the 80's by
the celebrated Aubry-Andr\'e (AA) model \cite{aubry1980analyticity},
has been renewed by the possibility of engineering QP modulations
in arrays of trapped atoms, cavity polaritons, and photonic lattices
\cite{lahini2009observation,tanese2014fractal,singh2015fibonacci,kohlert2019observation,yao2019critical,liu2021anomalous,An2021}
and by the emergence of moir\'e systems, such as twisted bilayer graphene
\cite{gonccalves2020incommensurability,wilson2020disorder}.

In the presence of interactions, the effects of the interplay between
quasiperiodicity and strong interactions is a subject under active
investigation \cite{vu2021moire,gonccalves2024incommensurability,gonccalves2024short}.
For one, it is yet unclear if electron-electron interactions and quasiperiodicity
combined can explain the physics of twisted bilayer graphene \cite{cao2018correlated,cao2018unconventional,gonccalves2020incommensurability}.

In interacting bosonic systems, the Bose condensed state may also
be affected by QP modulations \cite{sanchez2005bose,eksioglu2004matter,roati2008anderson,modugno2009exponential}.
Its effects can even be observed in one dimension, where a macroscopic
occupation of the condensed state is not possible and, instead, the
superfluid phase is characterized by an occupation of the most populated
state that grows as the square root of the total number of bosons
\cite{rigol2005hard}. This so-called quasicondensed state can be
destroyed by a QP perturbation yielding a compressible insulating
phase dubbed Bose glass \cite{yao2019critical,yao2020lieb,lellouch2014localization,damski2003atomic,d2014observation,roth2003phase,fallani2007ultracold}.
In the strongly-coupling limit, where repulsive on-site interactions
render bosons effectively hard-core particles, these effects have
been well-established thanks to the Jordan-Wigner (JW) mapping onto
non-interacting fermions \cite{rigol2004universal}. As a result,
a numerically exact analysis may be conducted for relatively large
system sizes \cite{ribeiro2013strongly,lellouch2014localization,wang2021many}.
This has permitted to show that, when submitted to a QP potential,
hard-core boson (HCB) lattices, host quasicondensed, Mott insulating
or Bose glass phases, depending on the location of the chemical potential
$\mu$ \cite{ribeiro2013strongly}. If $\mu$ lies in a spectral region
where JW single-particle states are delocalized (localized), the system
is a quasicondensate (Bose glass), and the fraction of particles in
the most occupied state, $\lambda_{0}$, behaves as $N_{b}^{1/2}$
$\left(N_{b}^{0}\right)$, with $N_{b}$ the total number of bosons.
If $\mu$ lies in a spectral gap the system is a Mott insulator with
$\lambda_{0}\sim N_{b}^{0}$. The behavior of HCB in the AA model
at criticality, including quantum dynamics analysis, was studied by
He et al. \cite{he2012noise,he2013single} and Gramsch et al. \cite{gramsch2012quenches}.
Nevertheless, the extension of such analysis to generalized AA models
and the study of the multifractal localization properties of HCB critical
states remains open.

\begin{figure}[h]
\centering{}\includegraphics[scale=0.42]{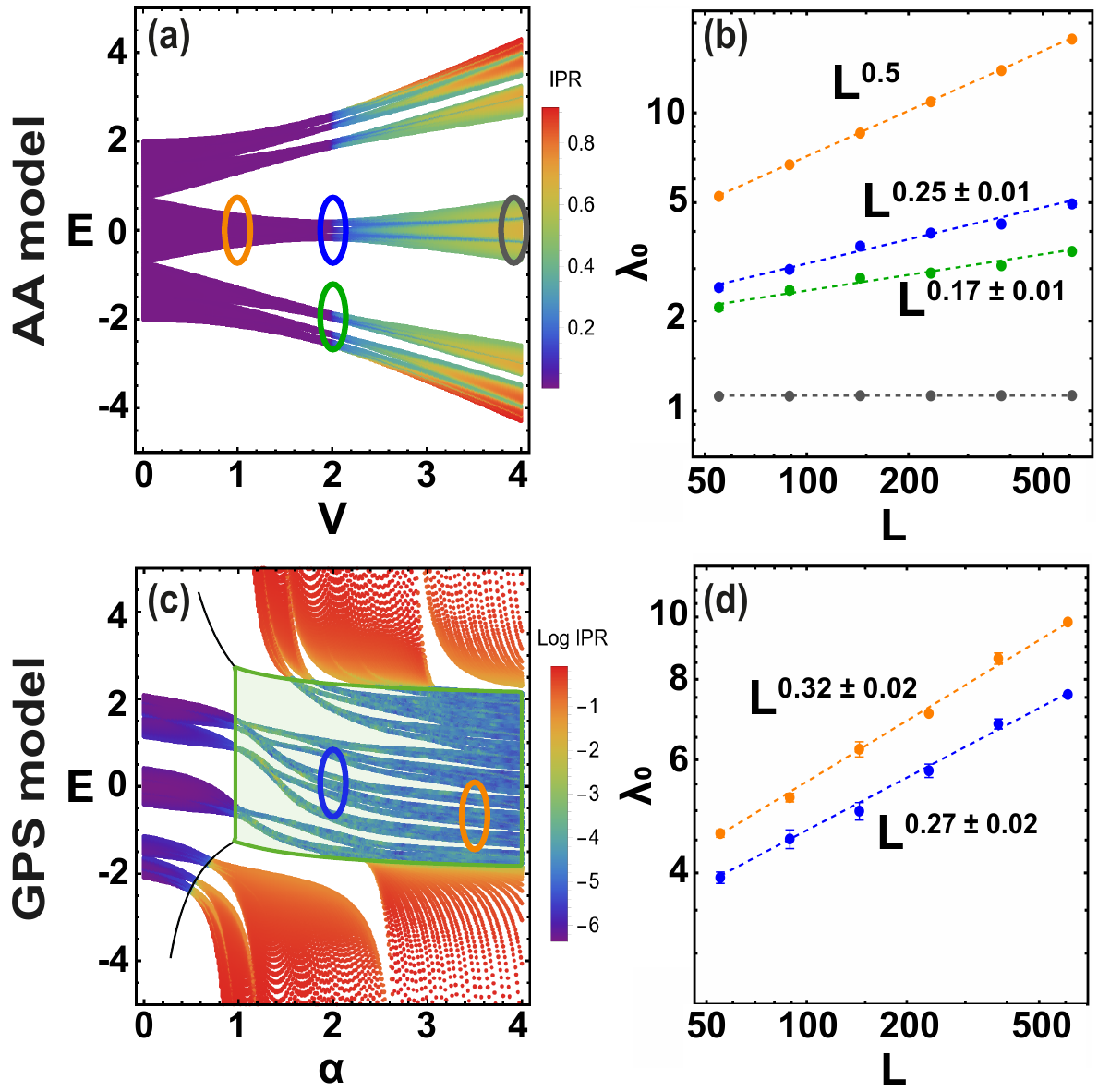}\caption{\textbf{AA model.} (a) Fermionic IPR for the single-particle eigenstates
as a function of the quasiperiodic potential strength, $V,$ and energy,
$E$, for system size $L=610$. (b) Scaling of the natural orbital's
highest eigenvalue, $\lambda_{0}$, as a function of $L$ for delocalized
($V=1$, orange), critical ($V=2$, blue), and localized ($V=4,$
gray) states at filling fraction $\nu=1/2$.\textbf{ }We also plot
the scaling of critical states for $V=2$ at $\nu=0.29$ (green) to
show that Bose superfluidity occurs for other filling fractions than\textbf{
$\nu=1/2$. GPS model.} (c) IPR of the JW fermions as a function of
$\alpha$ and $E$, computed for $V=0.75$ and $L=610$. Black curves
indicate extended-localized transitions. Critical phase is shaded
in green. (d) Scaling of $\lambda_{0}$ with $L$ for $\alpha=2$
at $\nu=0.48$ ($\alpha=3.5$ at $\nu=0.368$), where blue (orange)
points are numerical results fitted by the dashed lines. Fractal quasicondensation
is revealed by the sub-linear scalings $\lambda_{0}\propto L^{\gamma}$
with $0<\gamma<1/2$. Results in (b,d) are averaged over 10 random
configurations of $\theta$ and $\phi$. Error bars are obtained from
the standard deviation of the configurational average. \protect\label{fig:fig1}}
\end{figure}

Here, we investigate the fate of the quasicondensed state when the
chemical potential lies within a spectral region of fractal single-particle
eigenstates arising at localization-delocalization transitions or
in extended phase-space regions of critical states \cite{ganeshan2015nearest,an2021interactions,liu2021anomalous}.
We show that critical 1D HCB are fractal quasicondensates, characterized
by fractional occupation $\lambda_{0}\sim N_{b}^{\gamma}$, with $0<\gamma<1/2$,
and that the quasicondensed state exhibits multifractal localization
properties. In the critical regime, the scaling exponent, $\gamma$,
was found to be non-universal. To illustrate our findings we consider
the AA model at criticality, and the Ganeshan-Pixley-Sarma (GPS) model
\cite{ganeshan2015nearest} having anomalous mobility edges. Some
of these results are summarized in Fig.$\,$\ref{fig:fig1}, where
we also show the single-particle inverse participation ratio (IPR)
throughout the phase diagram of the AA and GPS models.

In the reminder of this article, we present the models and detail
our analysis of the occupations and of the properties of the fractal
quasicondensed state, we discuss our findings, and how our results
may be observed experimentally. Additional data corroborating our
conclusions and a discussion on particular implementations are provided
in the Appendices.

\section{Model and Methods}

We consider $N_{b}$ HCB on a 1D lattice with $L$ sites and periodic
boundary conditions. The Hamiltonian is given by:

\begin{equation}
H=-\sum_{n}(tb_{n}^{\dagger}b_{n+1}+\textrm{h.c.})+\sum_{n}V_{n}^{\gamma}b_{n}^{\dagger}b_{n},\label{eq:1}
\end{equation}
where $b_{n}\,(b_{n}^{\dagger})$ is the bosonic annihilation (creation)
operator at site $n=1,...,L$. The hard-core limit imposes the constrains
$b_{n}^{2}=b_{n}^{\dagger2}=0$ and imply the same-site anti-commutation
relation, $\left\{ b_{n},b_{n}^{\dagger}\right\} =1$, in addition
to the usual commutation relations, $\left[b_{n},b_{m}^{\dagger}\right]=0$,
for $n\neq m$. $t$ is the hopping integral and $V_{n}^{\gamma}$
is the on-site incommensurate potential specified below ($\gamma$
labels the two potentials considered). We apply twisted boundary conditions,
i.e. $b_{n}^{\dagger}=b_{n+L}^{\dagger}e^{i\theta}$, with phase twists
$\theta$, that can be easily implemented through the Peierls substitution,
$t\rightarrow te^{i\theta/L}$. Subsequent numerical results are presented
in units of $t$.

Concerning the QP potential, $V_{n}^{\gamma}$, the index $\gamma=\text{AA},\ \text{GPS}$
labels the two considered models, with potentials respectively given
by:

\begin{equation}
V_{n}^{\text{AA}}=V\cos(2\pi\tau n+\phi),\label{eq:3}
\end{equation}
\begin{equation}
V_{n}^{\text{GPS}}=\frac{V\cos(2\pi\tau n+\phi)}{1+\alpha\cos(2\pi\tau n+\phi)}.\label{eq:4}
\end{equation}
The parameters $(V,\ \alpha)$, the phase shift $\phi$, and $\tau$,
the ratio between the periods of the 1D lattice and the QP modulation,
fully characterize the potential. We take $\tau$ to be the inverse
of the Golden Ratio, $\tau=\varphi_{R}^{-1}=\left(\sqrt{5}-1\right)/2$.
For reducing finite-size effects, we consider rational approximants
given by the ratio of two successive Fibonacci numbers , $\tau_{j}=F_{j-1}/F_{j}$,
with $\tau=\tau_{\infty}$, and take $L=F_{j}$. We also average the
numerical results over random boundary twists, $\theta$, and shifts,
$\phi$, to further reduce finite-size effects.

For the AA model, transitions between delocalized ($0<V<2t$) and
localized ($V>2t$) states are energy independent and occur at $V_{c}=2t$,
as shown in Fig.$\,$\ref{fig:fig1}(a). For the GPS model, the mobility
edge is given by $E=\left(V\pm2t\right)/\alpha$ \cite{ganeshan2015nearest},
while the critical region is delimited by $|V-E\alpha|\leq|2t\alpha|\wedge|\alpha|\geq1$
\cite{liu2021anomalous}, as seen in Fig.$\,$\ref{fig:fig1}(c).

After the JW mapping, $b_{n}^{\dagger}=c_{n}^{\dagger}\prod_{\beta=1}^{n-1}e^{-i\pi c_{\beta}^{\dagger}c_{\beta}}$,
with $c_{n}(c_{n}^{\dagger})$ the fermionic annihilation (creation)
operator, the fermionic Hamiltonian is given by Eq.~$\eqref{eq:1}$,
replacing $b_{n}$ by $c_{n}$. The bosonic single-particle density
matrix (SPDM), $\rho_{nm}^{B}=\langle b_{m}^{\dagger}b_{n}\rangle$,
can be efficiently computed from its fermionic counterpart, $\rho_{nm}^{F}=\langle c_{m}^{\dagger}c_{n}\rangle$,
computed by evaluating matrix determinants \cite{rigol2005hard,rigol2004universal}.
Since the form of $H$ remains unchanged after the JW mapping, both
HCB and free fermions share the same energy spectrum. For diagonal
entries $\rho_{nn}^{B}=\rho_{nn}^{F}$, thus differences between fermionic
and bosonic single-particle density matrices are encoded in their
off-diagonal correlations.

For homogeneous systems at zero temperature, quasicondensation is
signaled by a divergence in the momentum distribution function at
zero momentum $n_{\kappa=0}\sim\sqrt{N_{b}}$ \cite{rigol2005hard}.
The generalization for spatially inhomogeneous systems amounts to
considering the highest eigenvalue of the bosonic SPDM, $\lambda_{0}$.
The eigenvectors of the SPDM, $\Phi_{j}^{n}$, are called \textit{natural
orbitals} \textit{(NO)}, i.e. \cite{penrose1956bose,leggett2001bose,rigol2004universal}:

\begin{equation}
\sum_{j}\rho_{ij}^{B}\Phi_{j}^{n}=\lambda_{n}\Phi_{i}^{n},\label{eq:2}
\end{equation}
with $\lambda_{0}\geq\lambda_{1}\geq\cdots$. The number of bosons
in the most occupied state scales with $N_{b}$, $\lambda_{0}\sim(N_{b})^{\gamma}$,
with $\gamma=0.5$ for quasicondensates associated with delocalized
states and $\gamma=0$ for Mott insulators and Bose Glasses.

In order to characterize phase transitions and analyze localization
properties of the wavefunctions, we use the inverse participation
ratio (IPR):

\begin{equation}
\text{IPR}[\psi]=\sum_{n}\left|\psi_{n}\right|^{4},\label{eq:5}
\end{equation}
where $\psi$ is the normalized fermionic wavefunction. The IPR shows
a power law scaling, $\text{IPR}[\psi]\sim L^{-\tau_{F}^{R}}$ ($R$
stands for real space), with $\tau_{F}^{R}=0$ for localized states,
$\tau_{F}^{R}=d$ for extended states ($d$ is the dimension) and
$\tau_{F}^{R}=D_{F}$ for multifractal states ($D_{F}$ is the fractal
dimension, obeying $0<D_{F}<d$) \cite{szabo2018non,fu2020magic,liu2021anomalous,he2013single}.
The scaling analysis of the $\tau_{F}^{R}$ exponent shows that, for
fermionic systems, the $\text{IPR}$ of both extended and multifractal
states tends to zero in the thermodynamic limit. Conversely, localization
in momentum space can be quantified by the momentum space IPR ($\text{IPR}_{K}$)
\cite{pixley2018weyl}:

\begin{equation}
\text{IPR}_{K}[\psi]=\sum_{k}|\tilde{\psi}_{k}|^{4},\label{eq:6}
\end{equation}
where $\tilde{\psi}_{k}$ are the Fourier coefficients of the wave
function. $\text{\ensuremath{\text{IPR}_{K}}}[\psi]\sim L^{-\tau_{F}^{K}}$,
with $\tau_{F}^{K}=d$ for localized and $\tau_{F}^{K}=0$ for ballistic
states. At criticality, both $\text{IPR}$ and the $\text{IPR}_{K}$
decrease with $L$. To study the localization properties of the bosonic
systems, we consider $\text{IPR\ensuremath{\left[\Phi^{n}\right]}}$
and the $\text{IPR}_{K}\left[\Phi^{n}\right]$, with $\Phi^{n}$ the
$n$-th NO.

\section{Results and Discussion}

We start with the AA model, defined by Eq.$\,$($\ref{eq:1}$) with
the on-site potential in Eq.$\,$($\ref{eq:3}$). For completeness,
we show the single-particle results in Fig.$\,$\ref{fig:fig1}(a),
where the well-known energy-independent localization transition at
$V_{c}=2$ is clearly seen in the IPR values of the JW fermions. Figure~\ref{fig:fig1}(b)
depicts the scaling of $\lambda_{0}$ with the system size at filling
fraction ($\nu=N_{b}/L$) $\nu=1/2$, for different values of $V$.
For $V<V_{c}$ ($V>V_{c}$), the scaling $\lambda_{0}\propto L^{\gamma}$with
$\gamma=1/2$ ($\gamma=0$) is recovered for the quasicondensed (Bose
glass) state. At criticality ($V=V_{c}$), the AA model behaves as
a Bose superfluid. For $\nu=1/2$ (blue curve), the exponent $\gamma\simeq0.25\pm0.01$
is obtained. Quasicondensation still occurs for $\nu$$\neq1/2$,
albeit with a smaller value for $\gamma$. The green curve in Fig.~\ref{fig:fig1}(b)
shows the results at criticality for $\nu=0.29$, where the corresponding
scaling exponent is given by $\gamma\simeq0.17\pm0.01$. Our results
are in accordance with He et al. \cite{he2012noise}.

Now we turn to the GPS model, given by Eq.$\,$$\eqref{eq:1}$ with
the incommensurate potential of Eq.$\,$$\eqref{eq:4}$. Figure~\ref{fig:fig1}$\,$(c)
shows the density plot of the single-particle fermionic IPR as a function
of the energy and of the parameter $\alpha$. For $\alpha<1$, only
extended-localized transitions are observed. Transitions between critical--extended
and critical--localized states occur at $\alpha=1$ and $\alpha>1$,
respectively.

Figure$\,$\ref{fig:fig1}(d) shows the scaling of $\lambda_{0}$,
for two values of the parameters $(\alpha,\nu)$ in the critical region
indicated in Fig.~\ref{fig:fig1}(c). As for the AA model, we find
$0<\gamma<1/2$, indicating the presence of an exotic quasicondensed
state. Specifically, we obtain $\gamma=0.27\pm0.02$ and $\gamma=0.32\pm0.02$
for $(\alpha,\nu)=(2,0.48)$ and $(\alpha,\nu)=(3.5,0.368)$, respectively.
We attribute the different scaling exponents to the parameter-dependent
fractal properties of the single-particle states \cite{PhysRevB.34.2041,PhysRevB.40.8225,szabo2018non}
which manifest in the bosonic NO through an exotic type of quasicondensation,
here dubbed fractal quasicondensation, with a tunable scaling exponent
$\gamma$. It should be noted that even for a model without energy-dependent
mobility edges, such as the AA model, such scaling exponent can be
tuned by means of the filling fraction.

\begin{figure}[h]
\includegraphics[scale=0.39]{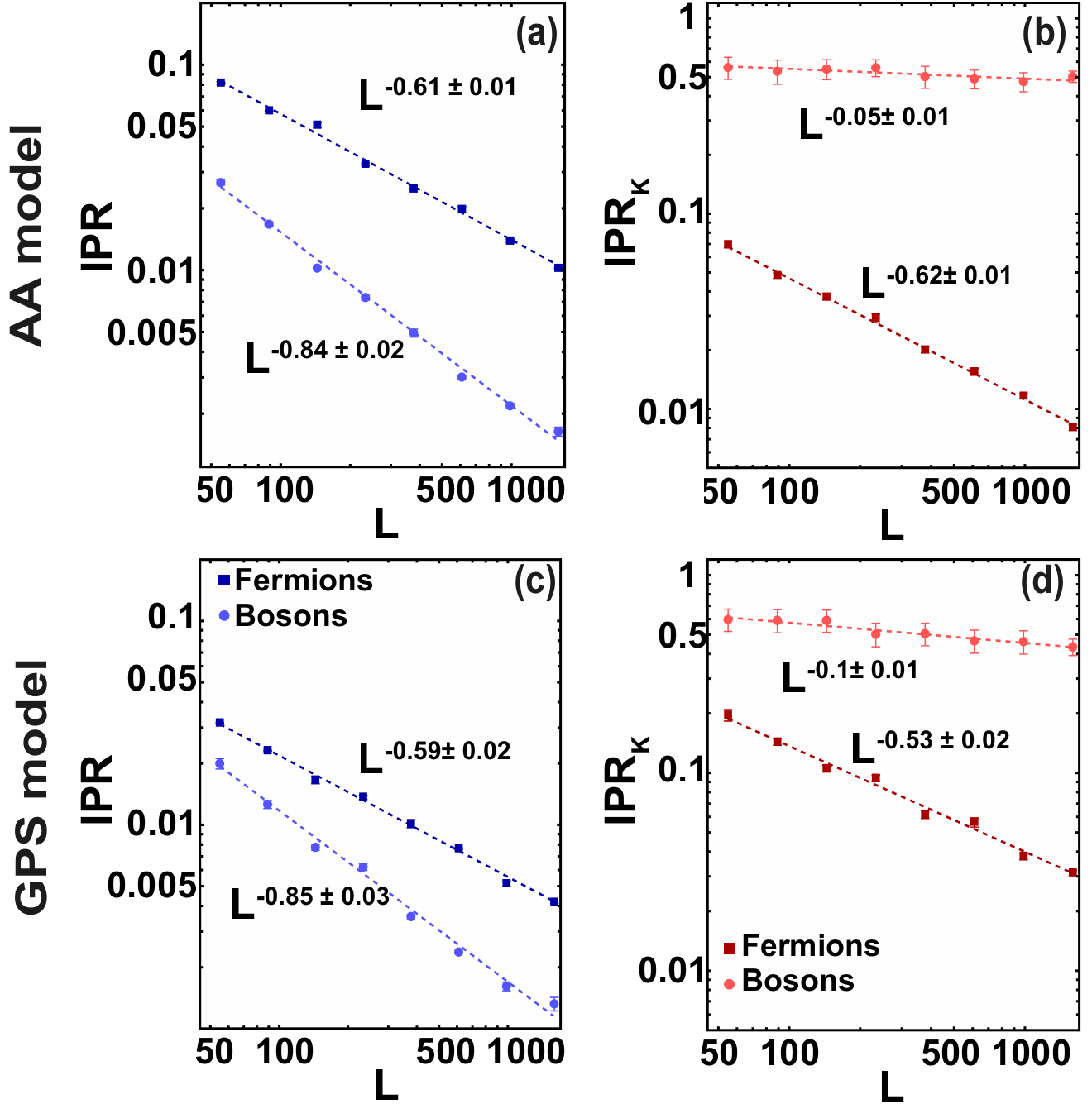}

\centering{}\caption{\textbf{AA model} ($V=2$). (a) IPR scaling of the fermionic eigenvectors
(natural orbitals), indicated by dark (light) blue points with error
bars, where the dark (light) red curve corresponds to the fitted model
$\textrm{IPR}\sim L^{-\tau_{F(B)}^{R}}$, with the exponent $\tau_{F}^{R}=0.61\pm0.01$
$(\tau_{B}^{R}=0.84\pm0.02)$. (b) IPR$_{K}$ scaling of the fermionic
eigenvectors (natural orbitals), indicated by dark (light) red points.
The IPR$_{K}$ of the fermionic eigenvectors scales as $\textrm{IPR}_{K}\sim L^{-\tau_{F}^{K}}$,
with $\tau_{F}^{K}=0.62\pm0.01$ $(\tau_{B}^{K}=0.05\pm0.01)$. \textbf{GPS
model} ($V=0.75,\ \alpha=3.5,\ \nu=0.368$). (c) IPR scaling of the
fermionic eigenvectors (natural orbitals), indicated by dark (light)
blue points, where the dark (light) curve is a fit yielding the the
exponent $\tau_{F}^{R}=059\pm0.02$ $(\tau_{B}^{R}=0.85\pm0.03)$.
(d) IPR$_{K}$ scaling of the fermionic eigenvectors (natural orbitals),
indicated by dark (light) red points, with $\tau_{F}^{K}=0.53\pm0.02$
$(\tau_{B}^{K}=0.1\pm0.01)$. Results are averaged over 30 random
configurations of $\theta$ and $\phi$.\protect\label{fig:fig2}}
\end{figure}

To investigate the fractal nature of the lowest NO (i.e. $\Phi^{n=0}$)
at criticality, we compute the scaling of their IPR and IPR$_{K}$,
as defined by Eqs.$\,$$\eqref{eq:5}$ and~$\eqref{eq:6}$, respectively.
The results are compared with the scalings of the fermionic single-particle
eigenstates at the same filling. This analysis is made for the AA
model at $V=2$ in Fig.$\,$\ref{fig:fig2}(a,b), and for the GPS
model in Fig.$\,$\ref{fig:fig2}(c,d).

Fig.$\,$\ref{fig:fig2}(a) shows that since $\ensuremath{0<\tau_{B}^{R}<1}$
($\tau_{B}^{R}=0.84\pm0.02$), the lowest NO is fractal albeit much
more delocalized in real space than its fermionic counterpart ($\tau_{F}=0.61\pm0.01$).
Concomitantly, the analysis of the $\text{IPR}_{K}$ in Fig.$\,$\ref{fig:fig2}(b)
yields $\tau_{F}^{K}=0.62\pm0.01$, $\tau_{B}^{K}=0.05\pm0.01$, and
indicates that the lowest NO is very localized in momentum space and
thus is much closer to a plane wave that the fermionic wave-function.
Together these results suggest a weak fractal nature of the NO. In
the following, we show that this is due to strong finite-size effects
and that the scaling exponents at criticality are fractal-like.

\begin{figure}[h]
\includegraphics[scale=0.57]{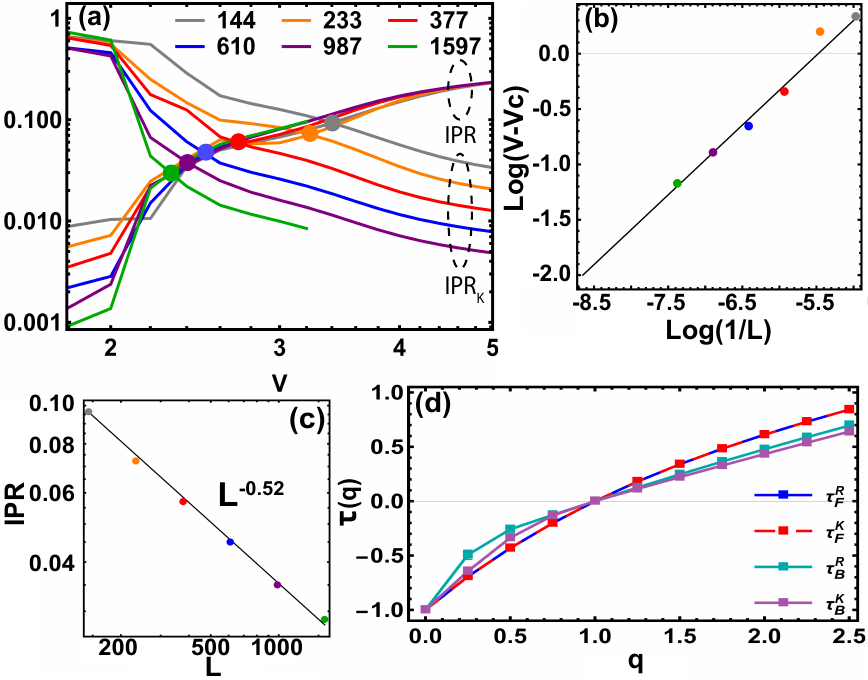}

\centering{}\caption{\textbf{AA model.} (a) IPR and IPR$_{K}$ of the NOs versus the on-site
potential ($V$) for various values of $L$. (b) Extrapolation of
the logarithm of the drift from the critical point versus $\log(1/L)$,
showing that the finite-size scaling is compatible with $V_{c}=2$.
(c) Scaling of IPR and IPR$_{K}$ for the lowest NO computed at the
crossing point of (a). (d) Scaling exponents $\tau_{F(B)}^{R/K}\left(q\right)$
versus the scaling parameter $q,$for free fermions (HCBs) in position,
$\tau_{F}^{R}$ ($\tau_{B}^{R}$), and momentum, $\tau_{F}^{K}$ ($\tau_{B}^{K}$),
spaces as an indicator of multifractality. \protect\label{fig:fig3}}
\end{figure}

Despite the disparities between the scaling exponents of the bosonic
and fermionic models, the localization transition occurs for a the
same critical potential $V_{c}=2$ in both models. This is analyzed
in Fig.~\ref{fig:fig3}(a) which displays the IPR (IPR$_{K}$) of
the NO as a function of the potential $V$, for different system sizes.
For small (high) $V$, the IPR (IPR$_{K}$) of the NO decreases with
$L$, whereas for high (small) $V$, the IPR (IPR$_{K}$) becomes
$L$-independent. As a result, at the transition point, these quantities
cross in the thermodynamic limit. However, Fig.$\,$\ref{fig:fig3}(b),
show that the finite $L$ crossing point has prominent finite-size
effects, which are much less severe in the fermionic case, but are
nevertheless compatible with $V_{c}=2$. At the same point, the critical
exponents of the IPR and IPR$_{K}$ estimated by extrapolating the
finite-size crossing point values, see Fig.$\,$\ref{fig:fig3}(c),
yields $\tau_{B}^{R}=\tau_{B}^{K}\sim0.52$ which clearly established
the fractal nature of the lowest NO. Thus, the exponents naively obtained
in Fig.$\,$\ref{fig:fig2}(a) and (b) that point to a weak fractal
nature of the NO are a consequence of the strong finite size effects.

To further study the fractal nature of the NO, in particular its multifractality
content, we consider the $q-$generalization of IPR \cite{fu2020magic,siebesma1987multifractal}
defined as:

\begin{equation}
\text{IPR}_{R/K}\left(q\right)=\sum_{r/k}|\psi_{r/k}|^{2q}\sim L^{-\tau^{R/K}\left(q\right)},\label{eq:7}
\end{equation}
and analyze its behavior at the the crossing point in Fig.$\,$\ref{fig:fig3}(d).
The onset of multifractality corresponds to a non-linearity dependence
of the exponet $\tau^{R/K}$ with $q$ \cite{fu2020magic,siebesma1987multifractal}.

For free fermions, we verify that $\tau_{F}^{R}\left(q\right)=\tau_{F}^{K}\left(q\right)$
for all $q$ as required by the self-duality of the AA model at $V=V_{c}$.
For HCBs, although the transition still arises for the same value
of $V$, the position and momentum space natural orbitals are no longer
self-dual. Interestingly, in this case, we observe strong finite size
effects at the critical point. Nevertheless, we still observe a non-linear
dependence of both $\tau_{B}^{R}\left(q\right)$ and $\tau_{B}^{K}\left(q\right)$
with $q$, signaling a the multifractal nature of the lowest NO.

\section{Experimental Implementation}

\begin{figure}[h]
\includegraphics[scale=0.4]{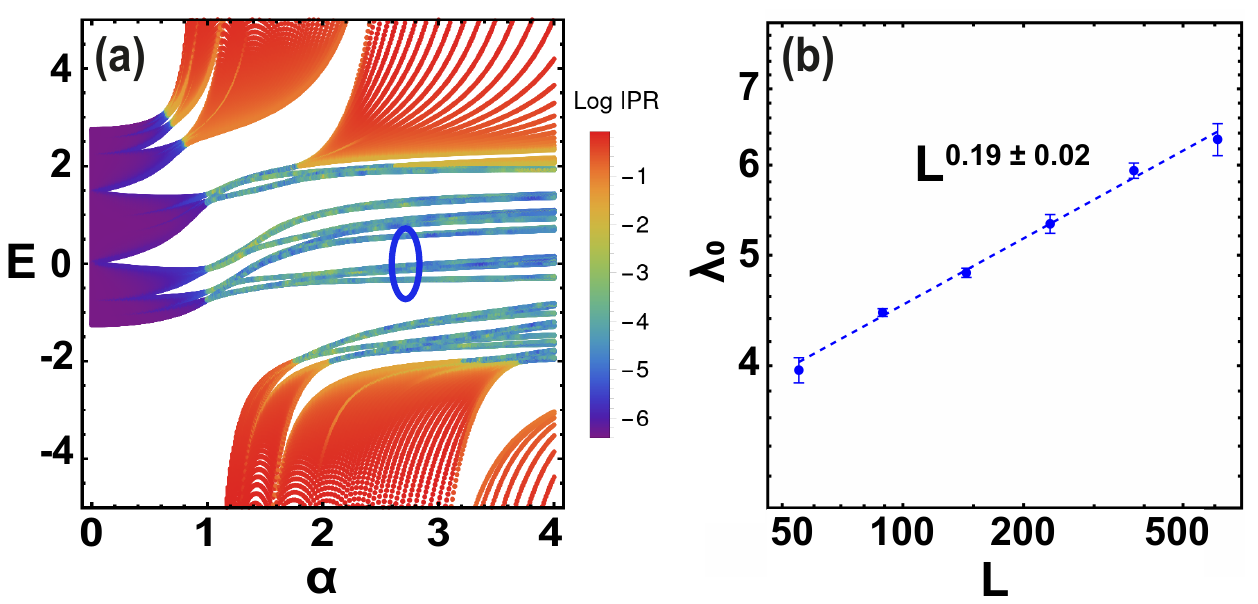}

\centering{}\caption{\textbf{Simplified version of the GPS model.} (a) Density plot of
the IPR as a function of the energy and of the on-site potential,
for $L=610$. (b) Scaling of the natural orbital's highest eigenvalue
($\lambda_{0}$) at the critical point, for $(\nu,\alpha)=(0.49,2.75)$.
(b) The quasicondensation revealed by the scaling of the occupation
of the lowest NO, $\gamma_{0}\sim L^{\gamma}$with $\gamma<1/2$.
\protect\label{fig:fig4}}
\end{figure}

In this section we argue that the mechanism of quasicondensation we
propose can be probed in current state-of-the-art experiments with
cold atomic gases. To avoid having to tune the system to a critical
point, we focus on the GPS model for which mutifractal states occupy
a finite region of the phase diagram. However, the implementation
of the QP potential described in Eq.~$\text{\eqref{eq:4}}$ may be
challenging, due its unbounded nature for $\alpha\geq1$. To overcome
this obstacle, we can resort to the\textit{ Floquet engineered Hamiltonian}
proposed by \cite{fishman1982chaos,grempel1984quantum,longhi2021maryland}
given by:

\begin{equation}
H=-\sum_{n}(tb_{n}^{\dagger}b_{n+1}+\textrm{h.c.})+\sum_{n}\frac{V}{1+\alpha\cos(2\pi\tau n+\phi)}b_{n}^{\dagger}b_{n}.\label{eq:8}
\end{equation}
As depicted in Fig.$\,$\ref{fig:fig4}(a), this simplified family
of models still possesses a region of critical states that is shown
to host quasicondensed states in Fig.$\,$\ref{fig:fig4}(b). These
models were shown to be physically realizable with conventional optical
lattice techniques by Liu et al.\cite{liu2021anomalous} for the
fermionic case. As in previous works \cite{greiner2002quantum,chin2010feshbach,will2012atom,wasak2014simple,schreiber2015observation,ray2015localization,deng2017tuning},
the adaptation of this method to HCB can be achieved by tuning the
Feshbach resonances as to make the interaction strength much higher
than any other energy scale, thus enforcing the hard-core constraint.
The critical nature of the states both in momentum and real space
can be inferred from time-of-flight experiments\cite{d2014observation}
and from direct measurements of the populations by absorption imaging\cite{An2021}.
Further details on the implementation of unbounded potentials using
Floquet engineered Hamiltonians are given in Appendix B.

\section{Conclusions}

In this article, we study the condensation of HCB in the presence
of quasiperiodic-induced critical states with multifractal wavefunctions.
We show that when the chemical potential is placed in regions where
the single-particle state are critical, quasicondensation acquires
exotic features. This regime, dubbed fractal quasicondensation, is
characterized by the growth of the occupation of the lowest natural
orbital, albeit with an exponent smaller than the value $\nu=1/2$
that is observed in one-dimension quasicondensates with ballistic
single-particle states. We analyze the real and momentum space structure
of the lowest NO and reveal its multifractal nature.

Finally, we propose how to implement such a generalized AA model that
hosts quasicondensed states. We hope our work sparks interest in the
observation and further exploration of these novel fractal quasicondensed
states, and leads to new insights in the interplay of strong interactions
and quasiperiodicity.

\section*{Acknoledgements}

The authors MG and PR acknowledge partial support from Funda\c{c}\~ao para
a Ci\^encia e Tecnologia (FCT-Portugal) through Grant No. UID/CTM/04540/2019. FR,
MG and PR acknowledge support by FCT through Grant No. UIDB/04540/2020. BA
and EVC acknowledge partial support from FCT- Portugal through Grant
No. UIDB/04650/2020. MG acknowledges further support from FCT-Portugal
through the Grant SFRH/BD/145152/2019. BA acknowledges further support
from FCT-Portugal through Grant No. CEECIND/02936/2017.

\section*{Appendix A: Additional Results for the GPS Model at Criticality\protect\label{sec:A}}

\renewcommand{\thefigure}{A\arabic{figure}}
\setcounter{figure}{0}

\begin{figure}[h]
\includegraphics[scale=0.4]{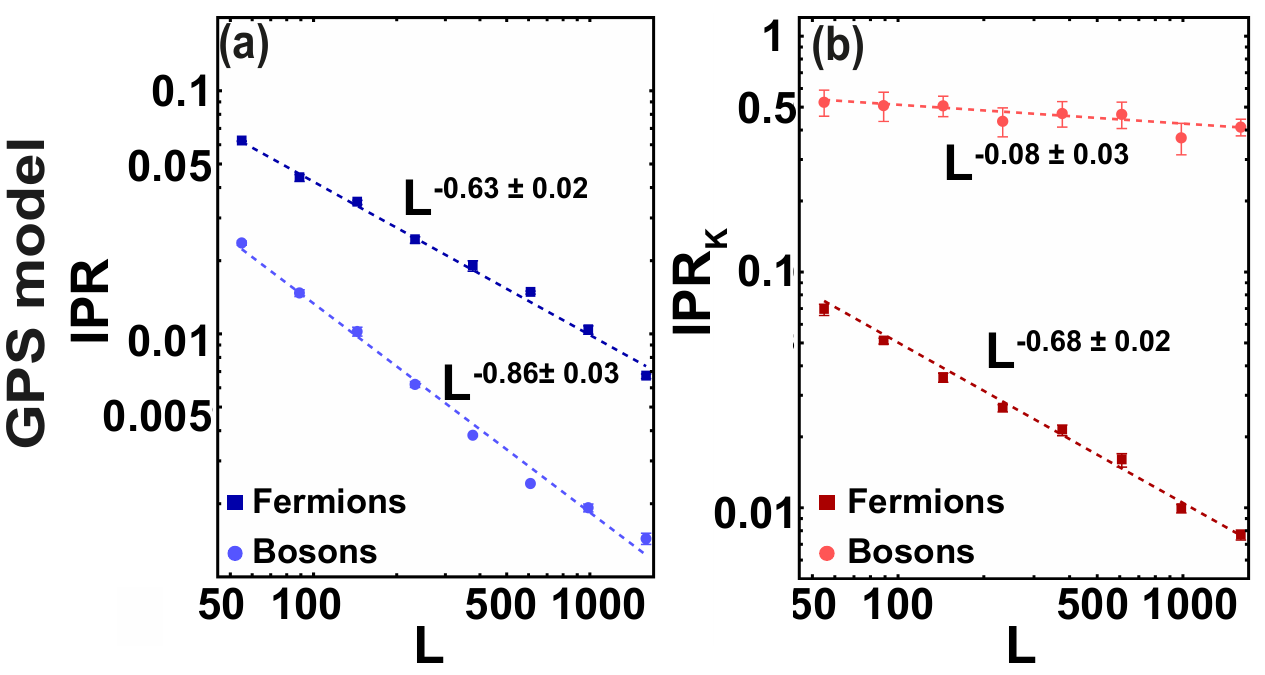}

\centering{}\caption{\textbf{GPS Model. }IPR and IPR$_{K}$ analysis inside the critical
phase region ($V=0.75,\ \alpha=2,\ \nu=0.48$). (a) IPR scaling of
the fermionic eigenvectors (bosonic natural orbitals), indicated by
dark (light) blue points, where the dark (light) curve corresponds
to the fitting, with the scaling exponent $\tau_{F}^{R}=0.63\pm0.02$
$(\tau_{B}^{R}=0.86\pm0.03)$. (b) IPR$_{K}$ scaling of the fermionic
eigenvectors (bosonic natural orbitals), indicated by dark (light)
red points, with the exponent$\tau_{F}^{K}=0.68\pm0.02$ $(\tau_{B}^{K}=0.08\pm0.03)$.
The results were averaged over 30 random configurations of twists
$\theta$ and phases $\phi$. \protect\label{fig:A1}}
\end{figure}

For completeness, we also show the results of the IPR scalings for
the fermionic vectors and for the natural orbitals (NOs) of another
phase-space state of the GPS model, with $V=0.75,\ \alpha=2,\ \nu=0.48$.
The results are similar to those obtained in the main article, depicted
in Figs.$\,$2(c,d). As we can see in Fig.$\,$\ref{fig:A1}, the
scaling exponent of the fermionic eigenvectors is less than one for
the IPR both in position and in momentum space. For the natural orbitals,
due to the finite-size effects discussed in the main text, the $\text{IPR}_{K}$
decreases very slowly, with a scaling exponent close to Fig.$\,$2(d).

\section*{Appendix B: Floquet Engineered Hamiltonians and the Implementation
of Unbounded Potentials\protect\label{sec:B}}

\renewcommand{\theequation}{B\arabic{equation}}
\setcounter{equation}{0}

In this section, we discuss in more details how to overcome problems
related with the physical implementation of the GPS model in order
to probe fractal quasicondensation. As mentioned in Section IV, \textit{Floquet
engineered Hamiltonians} are an effective way to overcome the potential
divergences associated with unbounded potentials such as those of
Eq. \ref{eq:4}. Here we show how to map the Floquet eigenvalue equation
onto the following tight-binding equation:

\begin{equation}
H=-t\sum_{x}(c_{x}^{\dagger}c_{x+1}+\textrm{h.c.})+\sum_{x}\frac{V}{1+\alpha\cos(2\pi\tau x+\phi)}c_{x}^{\dagger}c_{x}.\label{eq:S1}
\end{equation}
Although this Hamiltonian is different form that of the GPS model,
given in Eq. \ref{eq:4}, we show below it yields the same qualitative
features, namely an extended critical phase. To implement this Hamiltonian,
we follow the approach of Ref.~\cite{liu2021anomalous} starting
from a periodically-kicked quantum 1D system, described by the Schr\"odinger
equation (in units where $\hbar=1$):

\begin{equation}
K(p)\sum_{n}\delta(t-nT)|X,t\rangle+V(x)|X,t\rangle=i\partial_{t}|X,t\rangle.\label{eq:S2}
\end{equation}
The advantage of kicking the kinetic term, instead of the more common
approach where the potential term is engineered, is that we can recover
Eq.~$\ref{eq:S1}$ in position space. This is particularly relevant
for ensuring the hard-core constraint. The evolution of the wavefunction
for one kick is given by:

\begin{equation}
e^{-i\sum_{x}V_{x}c_{x}^{\dagger}c_{x}}e^{-i\sum_{p}K_{p}c_{p}^{\dagger}c_{p}}|x,t=m\rangle=|x,t=m+1\rangle,\label{eq:S3}
\end{equation}
where $V(x)=\sum_{x}V_{x}c_{x}^{\dagger}c_{x}$ and $K(p)=\sum_{p}K_{p}c_{p}^{\dagger}c_{p}$.
By defining $|X,t=m\rangle\equiv|X\rangle e^{-i\mu m}$, where $-\pi\leq\mu\leq\pi$
is the Floquet quasi-energy, we can rewrite Eq.~$\eqref{eq:S3}$
in terms of an eigenvalue problem:

\begin{equation}
e^{-i\sum_{x}V_{x}c_{x}^{\dagger}c_{x}}e^{-i\sum_{p}K_{p}c_{p}^{\dagger}c_{p}}|X\rangle=e^{-i\mu}|X\rangle.\label{eq:S4}
\end{equation}
The next step is to introduce the auxiliary operator, $W(p)\equiv\tan\left(\frac{\sum_{p}K_{p}c_{p}^{\dagger}c_{p}}{2}\right)$,
which gives:

\begin{equation}
e^{-i\sum_{p}K_{p}c_{p}^{\dagger}c_{p}}=\frac{1-iW(p)}{1+iWp)}.\label{eq:S5}
\end{equation}
We also define:

\begin{equation}
|X\rangle\equiv[1+iW(p)]|x\rangle.\label{eq:S6}
\end{equation}
Eqs.~$\eqref{eq:S5}$ and~$\eqref{eq:S6}$ in Eq.~$\eqref{eq:S4}$
yields:

\begin{equation}
[1+iW(p)]|x\rangle=e^{i\mu-i\sum_{x}V_{x}c_{x}^{\dagger}c_{x}}[1-iW(p)]|x\rangle,\label{eq:S7}
\end{equation}
which can be rewritten as:

\begin{equation}
W(p)|x\rangle=S_{x}|x\rangle,\label{eq:S8}
\end{equation}
where $\frac{1+iS_{x}}{1-iS_{x}}=e^{i\mu-i\sum_{x}V_{x}c_{x}^{\dagger}c_{x}}$.
Also:

\begin{equation}
W(p)=\sum_{p}W_{p}c_{p}^{\dagger}c_{p}|p\rangle\langle p|=\sum_{x,x'}|x\rangle W_{x-x'}\langle x'|,\label{eq:S9}
\end{equation}
and from Eq.~$\eqref{eq:S9}$ in~$\eqref{eq:S8}$:

\begin{equation}
\sum_{x,x'}|x\rangle W_{x-x'}\langle x'|x\rangle=S_{x}|x\rangle.\label{eq:S10}
\end{equation}

Now we show how to recover the tight-binding model of Eq.~$\eqref{eq:S1}$,
starting from the left-hand side of Eq.~$\text{\eqref{eq:S10}}$,
from which we are going to derive the hopping terms. Since $K(p)=\sum_{p}K_{p}c_{p}^{\dagger}c_{p}$
is diagonal in Bloch basis, with energies $\varepsilon_{p}=-2t\cos(p)$,
we can write the auxiliary operator as:

\begin{equation}
W(p)=\tan\left(\frac{\varepsilon_{p}}{2}\right)\sim\frac{\varepsilon_{p}}{2}=-t\cos(p),\label{eq:S11}
\end{equation}
since the kick time is much smaller than the inter-kick time, $\tau\ll T$.
Considering only first neighbors, Eq.~$\text{\eqref{eq:S10}}$ reads:

\begin{equation}
-t|x-1\rangle-t|x+1\rangle=S_{x}|x\rangle.\label{eq:S12}
\end{equation}

The right-hand side of Eq.~$\eqref{eq:S8}$ will give us the QP potential
and the eigenvalues based on the Floquet quasi-energy. By defining
$E\equiv1/\tan(\mu/2)$ and rewriting $\frac{1+iS_{x}}{1-iS_{x}}=e^{i\mu-i\sum_{x}V_{x}c_{x}^{\dagger}c_{x}}$
we obtain

\begin{equation}
S_{x}=E-\sum_{x}\frac{V}{1+\frac{1}{E}\tan\left(\frac{V_{x}}{2}\right)}c_{x}^{\dagger}c_{x},\label{eq:S13}
\end{equation}
where $V\equiv\frac{1+E^{2}}{E}$.

To recover Eq.~$\text{\eqref{eq:S1}}$ we set $\ensuremath{V_{x}=2\arctan[A\cos(2\pi\tau x+\phi)]}$.
The QP potential in Eq.~$\text{\eqref{eq:S13}}$ becomes:

\begin{equation}
\sum_{x}\frac{V}{1+\alpha\cos(2\pi\tau x+\phi)}c_{x}^{\dagger}c_{x},\label{eq:S14}
\end{equation}
where $\alpha=A/E$, varying without divergences.

\bibliographystyle{IEEEtranS}
\bibliography{fractal_condensation_nov_11.bbl}

\end{document}